\documentclass[final]{aipproc}

\layoutstyle{6x9}

\usepackage{amssymb}

\begin{document}

\title{Low temperature mean opacities for the carbon-rich regime}


\author{Michael T. Lederer}{
  address={
  	Department of Astronomy, University of Vienna,
		T\"urkenschanzstra\ss e 17,
		A-1180 Wien, Austria
		}
}

\author{Bernhard Aringer}{
  address={
  	Department of Astronomy, University of Vienna,
		T\"urkenschanzstra\ss e 17,
		A-1180 Wien, Austria
		}
		,
		altaddress={
		Dipartimento di Astronomia, Universit\`a di Padova,
		Vicolo dell'Osservatorio 3, 35122 Padova, Italy
		}
}

\begin{abstract}
Asymptotic Giant Branch (AGB) stars undergo a change in their chemical composition during their evolution. This in turn leads to an alteration of the radiative opacities, especially in the cool layers of the envelope and the atmosphere, where molecules are the dominant opacity sources. A key parameter in this respect is the number ratio of carbon to oxygen atoms (C/O). In terms of low temperature mean opacities, a variation of this parameter usually cannot be followed in stellar evolution models, because up to now tabulated values were only available for scaled solar metal mixtures (with C/O $\simeq0.5$). We thus present a set of newly generated tables containing Rosseland mean opacity coefficients covering both the oxygen-rich (C/O $<1$) and the carbon-rich (C/O $>1$) regime. We compare our values to existing tabular data and investigate the relevant molecular opacity contributors.
\end{abstract}

\maketitle


\section{Introduction}

Low and intermediate mass stars, depending on their actual mass and metallicity, change their surface chemical composition in the AGB phase of their evolution due to a phenomenon that is usually dubbed "third dredge-up" (see \citealp{1999ARA&A..37..239B} for a detailed review). This process denominates a series of so-called thermal pulses (TP), where at the end of each pulse the convective envelope penetrates the sites of nucleosynthesis and the freshly produced elements are brought to the surface of the star. Of particular concern are the ashes of He burning, which is mainly $^{12}$C. From the several reasons of why the dredge-up of carbon is of great importance for the AGB evolution, the one being particularly considered here is the change in opacity due to the carbon enhancement. Radiative opacities are provided by the OPAL collaboration \citep{1993npsp.conf..221R} as well as the Opacity Project \citep{2005MNRAS.362L...1S}. In both cases the effects of single element abundance variations are covered by the data. Unfortunately they do not include the most important molecular opacity sources but are basically limited to the atomic contributions. In the cool layers (i.e., temperatures lower than about $4000\,\rm{K}$) of the convective envelope and the atmosphere of late-type stars, however, molecules become the dominant opacity source.

For the temperature range below $10^4\,\rm{K}$, a common choice are the tabulated
Rosseland opacities from \citet{2005ApJ...623..585F} [F05]. This compilation contains only tables for a scaled solar metal composition (plus tables with varied abundances of the
alpha elements). As outlined above, the chemical composition of an AGB star changes during its evolution. The amount of molecules formed for certain species depends, beside the local thermodynamic conditions, crucially on the number ratio of carbon to oxygen atoms (C/O) which is about $0.5$ for solar abundances (the exact value depends on the adopted abundance prescription). Despite the fact that the use of the F05 tables substantially underestimates the radiative opacity in the cool atmosphere of an evolved AGB star eventually enriched in C and N, they are commonly adopted by stellar modelers. \citet{2002A&A...387..507M} made a first step toward a correct description of the abundance changes in the calculation of opacity coefficients by estimating molecular concentrations through dissociation equilibrium calculations. These were combined with approximation formulae for the contribution of each molecular species to the Rosseland mean opacity. Although the results of this work definitively demonstrate the importance of a correct opacity treatment, its simplified approach suffers from some drawbacks, for instance the limited number of molecular species included.

In the following we present a set of new opacity tables that appropriately take into account possible overabundances of carbon and nitrogren with respect to a scaled solar metal mixture. In this way one can in terms of radiative opacity correctly follow the evolution of stars undergoing third dredge-up. Additionally, the consequences of a slow deep mixing process that is generally considered responsible for the partial conversion of carbon into nitrogen in low-mass AGB stars can be investigated.

\section{Opacity tables}

\begin{figure}\label{fig:mtl-comparison}
  \includegraphics[width=1.0\textwidth]{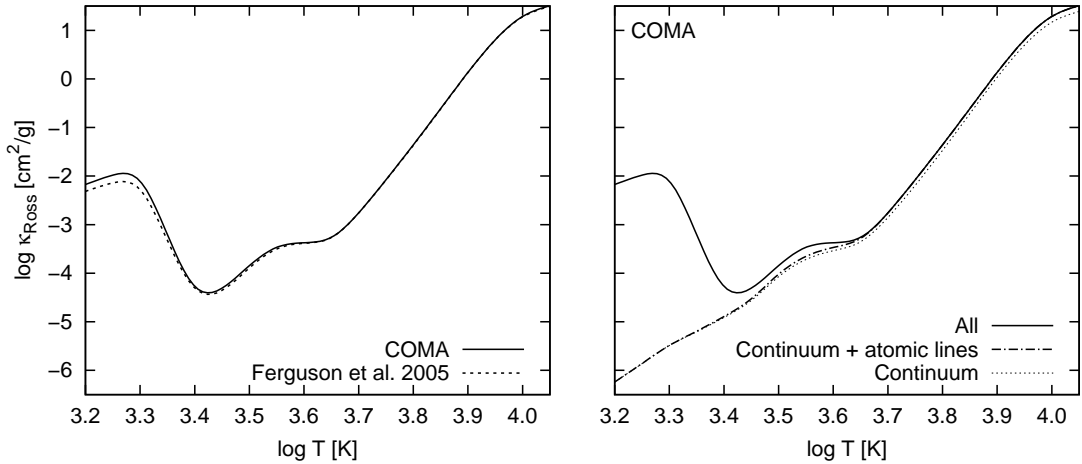}
\caption{
Left panel: Comparison of the Rosseland mean opacity calculated with COMA to the values of F05 for solar metallicity ($X=0.7$, $Z=0.02$) at $\log R=-3.0$. Metal abundances are taken from \citet{2003ApJ...591.1220L}. The data match for high temperatures but do differ at lower temperatures which is mostly due to the different set of molecular line lists used in the particular calculations. 
\newline
Right panel: Contribution of continuum, atomic (evaluated by omitting all line data and only molecular line data in the calculations, respectively) and molecular sources to the total Rosseland mean opacity (abundances and parameters as given above). Molecules are the dominating opacity source at low temperatures, whereas at higher temperatures the opacity is almost all due to continuum sources. The contribution of atomic line opacity is relatively minor but non-negligible.}
\end{figure}

The new opacity tables have been calculated by means of the COMA code from \citet{Aringer2000}.
Beside continuous opacity sources and atomic line opacities, the latter ones being derived from VALD \citep{2000BaltA...9..590K}, we included data for a total number of 20 molecules in our calculations. The list of molecules given in \citet{2007ApJ...667..489C} including references to corresponding thermodynamic data has recently been extended: we added data for the molecules ZrO (Bertrand Plez, priv. comm.; this list was used for the generation of a model atmosphere grid for S stars by \citealp{2003IAUS..210P..A2P}), YO (John Littleton, priv. comm.), FeH \citep{2003ApJ...594..651D} and CrH \citep{BauschlicherRam2001}. Grain opacities have not been included in our calculations, however, the supposed scope of the produced data are stellar evolution models that usually do not attain such low temperatures.

In order to ease the integration of our tables into existing stellar evolutions codes we adopted the format of the \citet{1994ApJ...437..879A} tables. In more detail, we tabulated values for the logarithm of the Rosseland mean opacity (in $\rm{cm}^2\,\rm{g}^{-1}$) as a function of $\log T$ and $\log R$\footnote{$R\equiv\rho/T_6^3$ where $[\rho]=\rm{g}\,\rm{cm}^{-3}$ and $T_6$ is the temperature in millions of Kelvin.}. We cover a range in temperature from $\log T=3.2$ to $\log T=4.05$ with steps of $\Delta \log T=0.05$, and in the $\log R$ dimension from $-7$ to $\log R=1$ with steps of $\Delta \log R=0.5$.

Up to now, tables have been calculated for metallicities of $Z=1\times 10^{-4}$, $1\times 10^{-3}$, $3\times 10^{-3}$, $6\times 10^{-3}$ and $Z=Z_\odot$. Starting from a given scaled solar metal composition --- the element abundances and the isotopic composition were taken from \citet{2003ApJ...591.1220L} --- with the overall metallicity as given above, we increased the mass fractions of $^{12}$C and $^{14}$N by factors depending on the initial metallicity. All other isotopes were left untouched. This results in a general increase of the metallicity, thus we reduced the mass fraction of $^{4}$He in turn (in consistency with the OPAL approach). For both $^{12}$C and $^{14}$N we applied 5 enhancement factors for three different values of $X$ ($0.5$, $0.7$, $0.8$), whereby we get 75 tables per initial metallicity. Variations in the mass fraction of $^{16}$O were also investigated. Compared to the effects of the carbon enhancement, this contribution was found to be negligible. For this reason, alterations of the O abundance were not included in our current set of tables. However, we plan to assess the implications of varying the alpha element abundances as a whole.

Before the generation of our set of tables we compared the results to data from \citet{2005ApJ...623..585F}. In Fig.~\ref{fig:mtl-comparison} we show an example for the solar metallicity case ($X=0.7$, $Z=0.02$) at $\log R=-3$. The agreement for temperatures higher than $\log T \gtrsim 3.6$ is very good, whereas the results differ somewhat for the lower temperatures. The reason for the discrepancy is partly due to the different values adopted for the microturbulent velocity (COMA: $\xi=2.5\,\rm{km}\,\rm{s}^{-1}$, F05: $\xi=2.0\,\rm{km}\,\rm{s}^{-1}$) in the evaluation of the line opacities, but the major part can be ascribed to the differing molecular line dataset. In the same figure we indicate the contributions of continuous, atomic and molecular opacity to the mean opacity coefficient. Molecules are the dominant opacity source at low temperatures,  continuous opacities are relevant at high temperatures. Atomic line opacities play a comparably minor role but have to be taken into account to arrive at accurate results.

\subsection{Molecular contribution}

\begin{figure}[!t]\label{fig:mtl-oxygenrichmolecules}
  \includegraphics[width=1.0\textwidth]{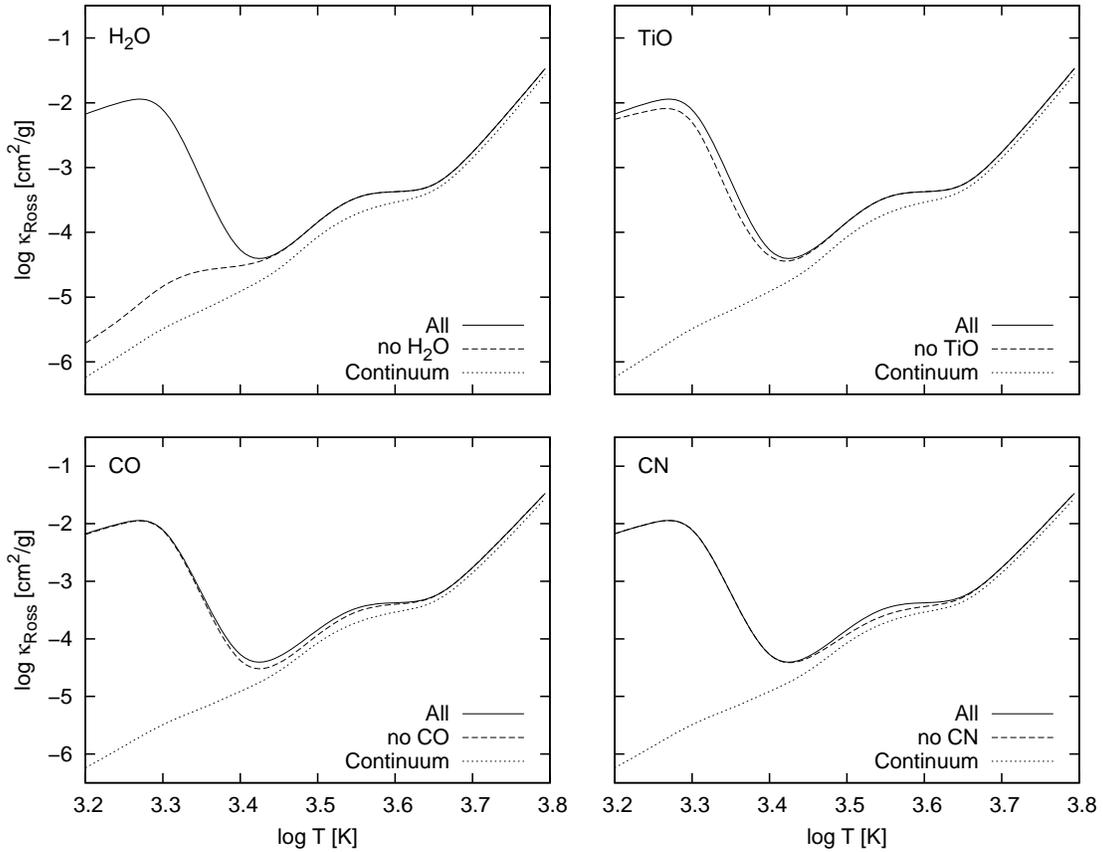}
\caption{
Molecules that contribute most to the Rosseland mean opacity in the oxygen-rich case (C/O $=0.50$).
At the lowest temperatures $\kappa_{\rm Ross}$ is completely dominated by the water molecule (H$_2$O). In this region, titanium oxide (TiO) also contributes significantly. At somewhat higher temperatures, CO and CN are important opacity sources. This is true in both the oxygen-rich and the carbon-rich regime. The solar metallicity case ($X=0.7$, $Z=0.02$) is shown at a value of $\log R=-3.0$. "All" denotes the calculation with the full molecular dataset, "no \ldots" means that the indicated molecule was omitted in the calculation.
}
\end{figure}

\begin{figure}[!t]\label{fig:mtl-carbonrichmolecules}
  \includegraphics[width=0.98\textwidth]{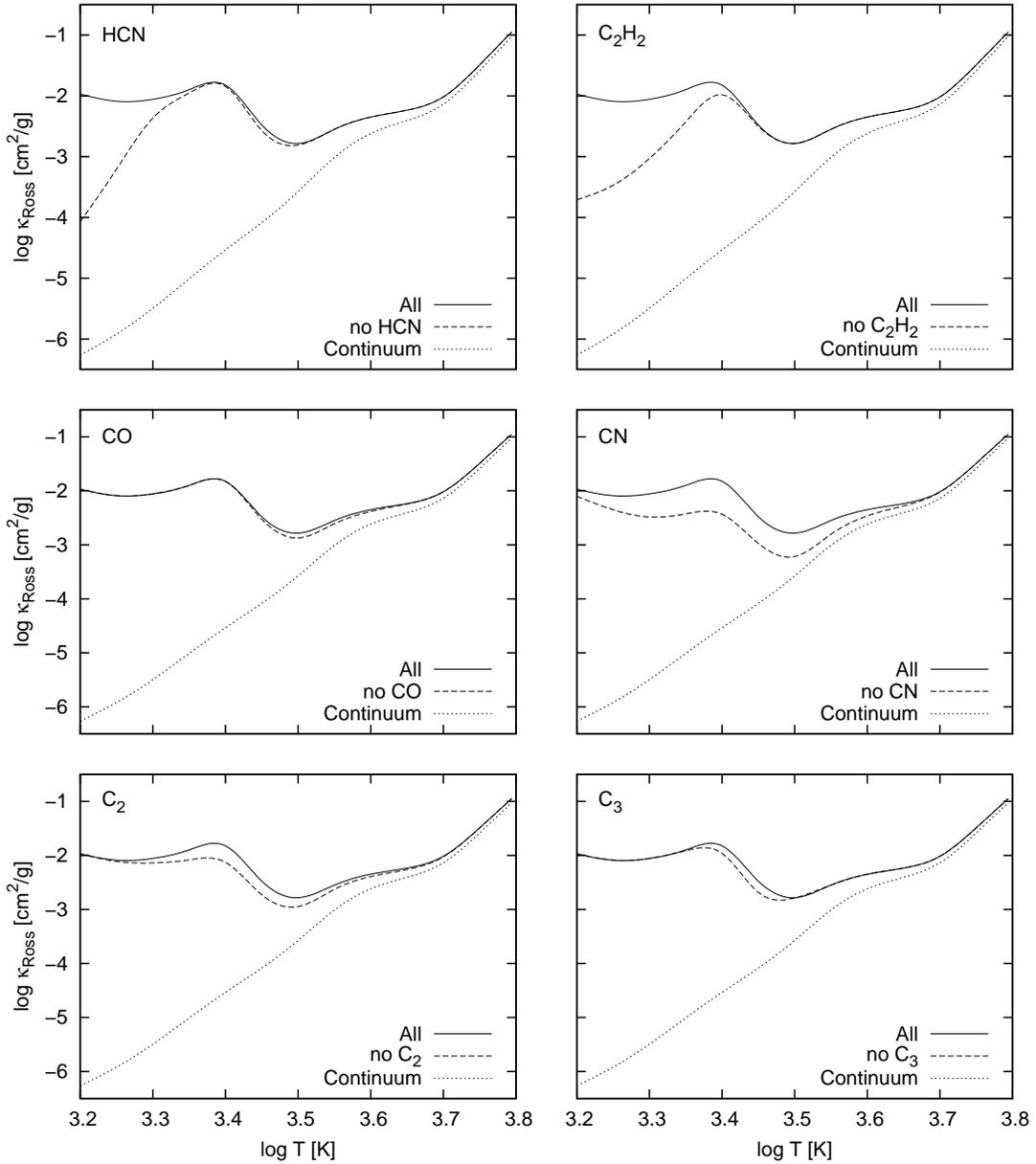}
\caption{
Molecules that contribute most to the Rosseland mean opacity in the carbon-rich case (C/O $=2.00$). C$_2$H$_2$ and HCN deliver the bulk of the opacity at low temperatures. The contribution of CO is a little bit lower than in the oxygen-rich case, whereas CN contributes considerably more to $\kappa_{\rm Ross}$ over a wider temperature range. C$_2$ and C$_3$ become important especially for higher C/O ratios and cause the peaked structure around $\log T=3.4$ in the plots shown here. Cf. also Fig.~\ref{fig:mtl-oxygenrichmolecules}.
}
\end{figure}

For the evaluation of the molecular contribution to the Rosseland mean opacity we calculated tables omitting the respective molecule under consideration and compared them to calculations with the full set of molecular data. Due to the non-linear character of the Rosseland mean one can only estimate the contribution of a single molecule in this indirect way. 

In Fig.~\ref{fig:mtl-oxygenrichmolecules} we identify H$_2$O, TiO, CN and CO to be the main constituents of $\kappa_{\rm Ross}$ in the oxygen-rich case. Other molecules that should be taken into account are VO, SiO and OH. For the value of $\log R=-3$ shown in Fig.~\ref{fig:mtl-oxygenrichmolecules}, VO contributes at $\log T\simeq3.3$, SiO and OH  are relevant around $\log T=3.4$. A calculation based on these seven molecules results in opacity coefficients that only deviate by a few percent from a calculation using the full dataset.

In the carbon-rich case, again CO and CN are important opacity sources. The largest component of $\kappa_{\rm Ross}$ at lower temperatures is however generated by the polyatomic molecules C$_2$H$_2$ and HCN. C$_2$ and C$_3$ are relevant at slightly higher temperatures and give rise to a peak in the opacity around $\log T\simeq3.4$ for increasing C/O ratios.

\begin{figure}[!t]\label{fig:mtl-transition}
  \includegraphics[width=1.0\textwidth]{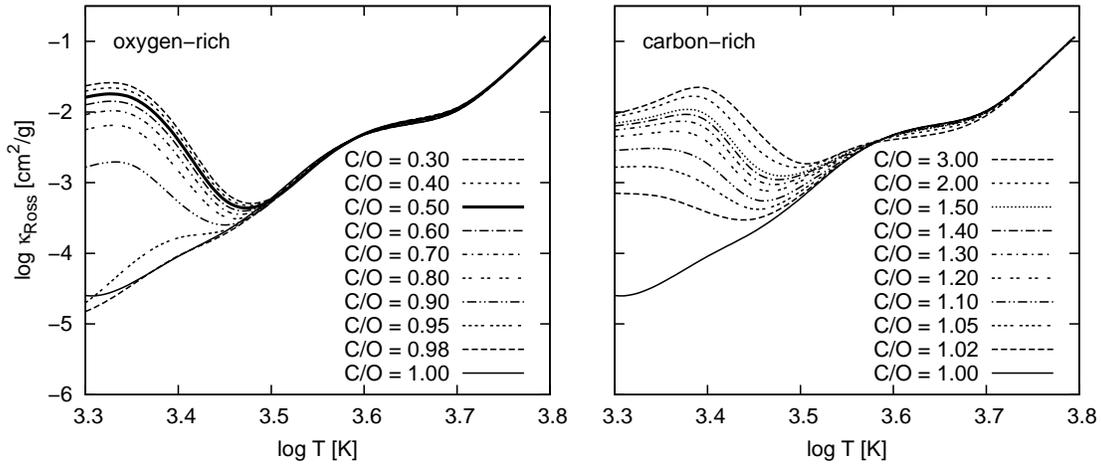}
\caption{
Changes of the mean opacity coefficient for an altered carbon-to-oxygen ratio. The solar value of C/O $=0.50$ is indicated by the thick solid line. In the oxygen-rich regime (C/O $<1$, left panel), an increasing C/O ratio causes the opacity at low temperatures to drop as the number of oxygen atoms that are free to form other molecules than CO decreases. After the transition to the carbon-rich regime, C/O $>1$, right panel) the opacity rises again as more and more carbon-bearing molecules are formed. The parameters for the plots are $X=0.7$, $Z=0.02$ and $\log R=-1.5$. Note that for this plot the mass fractions of all the metals have been rescaled after enhancing the amount of $^{12}$C to arrive at $Z=0.02$ again.
}
\end{figure}

\subsection{C/O variations}

The reason why the number ratio of carbon to oxygen atoms C/O is the decisive quantitiy for molecular opacities is the large dissociation energy of the CO molecule. For C/O $<1$ almost all carbon atoms are bound in CO while the oxygen atoms in excess are free to form other molecules as for example H$_2$O and TiO. As C/O increases, less oxygen-bearing molecules (apart from CO) are formed which causes the mean opacity to drop (see Fig.~\ref{fig:mtl-transition}, left panel). Around C/O $=1$ the Rosseland mean opacity reaches a minimum at low temperatures. A further increase in C/O beyond $1$ (when the star becomes carbon-rich) induces the formation of carbon-bearing molecules such as C$_2$, C$_3$, C$_2$H$_2$, HCN etc. and the opacity rises again (right panel of Fig.~\ref{fig:mtl-transition}). It is thus crucial to follow the chemical changes in the envelopes of AGB stars as the dredge-up of carbon results in significantly different opacities not only when transiting from the oxygen-rich to the carbon-rich regime, but even within both regimes.

\begin{figure}[!t]\label{fig:mtl-showcase}
  \includegraphics[width=1.0\textwidth]{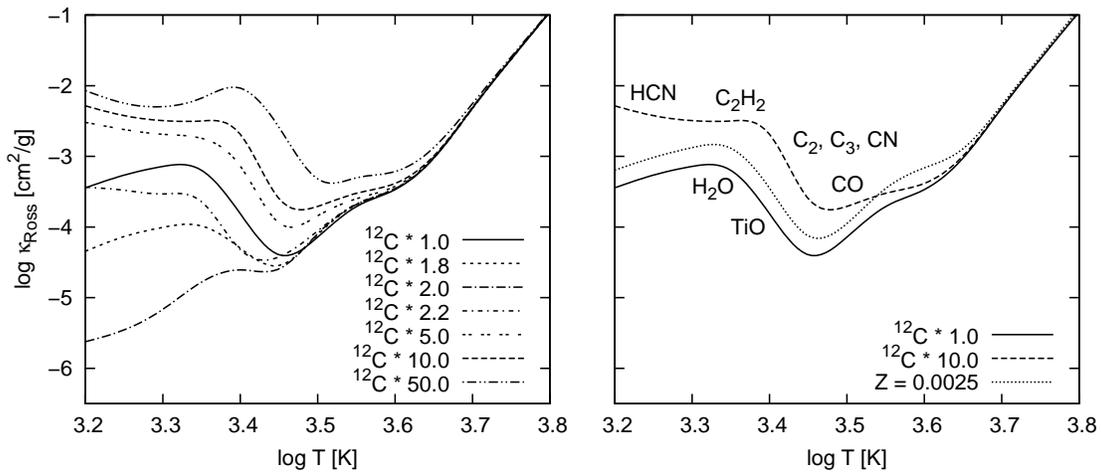}
\caption{
Left panel: Example of a set of opacity tables for $Z=1\times 10^{-3}$ (metal composition from to \citealp{2003ApJ...591.1220L}). The mass fraction of $^{12}$C was enhanced according to the factors indicated. This results, beside a higher C/O ratio, in an overall increase of $Z$. Thus the $^{4}$He mass fraction was reduced instead to fulfil $X+Y+Z=1$. Parameters for both plots in this figure are $X=0.7$ and $\log R=-1.5$ \newline
Right panel: Comparison of data obtained by only enhancing the $^{12}$C (dashed line) abundance with a table based on a scaled solar metal mixture with the same overall metallicity (dotted line). The mean opacity for a metal mixture with C/O $> 1$ cannot be approximated by a table for the oxygen-rich case with a higher metallicity. Both the shape of the curve (relevant molecules are indicated for both cases) as well as the order of magnitude differ significantly. 
}
\end{figure}

In Fig.~\ref{fig:mtl-showcase} we show an example for a set of opacity tables at a metallicity of $Z=1\times10^{-3}$. As described above (and in contrast to the data shown in Fig.~\ref{fig:mtl-transition}) no rescaling of the metals is done but the $^{4}$He mass fraction is reduced instead. When the star becomes carbon-rich, molecules do form that are more opaque than the ones in an oxygen-rich regime, which causes a steeper temperature gradient. A consequence is for instance a decrease of the effective temperature in stellar evolution models. The stellar radius increases, and the average mass-loss rate increases, thus eroding at a faster rate the envelope mass. As shown by \citet{2003PASA...20..389S}, changes of the core and the envelope masses affect all the fundamental properties of AGB stars, such as the TP strength or the total amount of mass that is dredged up.

Moreover, the right panel of the figure underlindes the fact that one cannot mimic the increase in the mass fraction of carbon in a metal mixture by simply scaling up all the metals starting from solar values. As different molecules contribute to the mean opacity it is qualitatively and quantitatively different in each case.

\section{Applications}
The new opacities were first applied in a stellar evolution model with a metallicity of $Z=1\times 10^{-4}$ calculated with the FRANEC code (the release described in \citealp{1998ApJ...502..737C}). To summarize the results briefly, the larger opacity coefficients in the carbon-rich regime imply cooler envelopes, larger mass-loss rates and thus shorter AGB lifetimes. Moreover, the variation of the surface composition and the global yields of low-mass AGB stars are affected. It turns out that the models computed were able to reproduce the physical properties of their observational counterparts. For an in-depth discussion we refer to the paper by \citet{2007ApJ...667..489C}. For the implications of the new opacity coefficient at other metallicities see the contribution of Cristallo et al., this volume. Details will be given in a series of forthcoming papers.

\section{Conclusions}

The new low temperature Rosseland mean opacities introduced here represent a step ahead in the modelling of low and intermediate mass stars. We want to point out that whenever temperatures lower than approximately $\log T=3.7$ are attained in a non-solar metal mixture (i.\,e.\  for C/O ratios other than $0.5$), the usually adopted scaled solar values for the mean opacity coefficients are inappropriate to describe the actual physics. The implications for stellar evolution could already be anticipated from the analysis of the data shown here, and the application of the tables in detailed model calculations confirmed our expectations. It appears that the inconsistency between surface temperatures and s-process enhancements arising from a comparison of the models to observations of carbon-enhanced metal-poor stars could be solved by using appropriate opacity coefficients (cf. the contribution of Cristallo et al., this volume). An interesting question would also be the interdependency of the new opacity coefficients with different mass loss prescriptions, as these issues are physically closely coupled.

The next step will be to establish a large grid of opacity tables and to collect them in a database. With the help of a complete dataset one will be able assess the impact of varying opacity coefficients on stellar evolution over a wide metallicity range.


\begin{theacknowledgments}
MTL and BA acknowledge funding by the Austrian Research Fund FWF (projects P-18171 and P-19503). MTL has been supported by the Austrian Academy of Sciences (DOC programme).
\end{theacknowledgments}



\begin{thebibliography}{15}

\bibitem[\protect\citeauthoryear{{Alexander} \& {Ferguson}}{{Alexander} \&
  {Ferguson}}{1994}]{1994ApJ...437..879A}
{Alexander} D.~R.,  {Ferguson} J.~W.,  1994, ApJ, 437, 879

\bibitem[\protect\citeauthoryear{{Aringer}}{{Aringer}}{2000}]{Aringer2000}
{Aringer} B.,  2000, PhD thesis, {University of Vienna}

\bibitem[\protect\citeauthoryear{{Bauschlicher} C.~W., {Ram}, {Bernath},
  {Parsons} \& {Galehouse}}{{Bauschlicher} et~al.}{2001}]{BauschlicherRam2001}
{Bauschlicher} C.~W. J.,  {Ram} R.~S.,  {Bernath} P.~F.,  {Parsons} C.~G.,
  {Galehouse} D.,  2001, J. Chem. Phys., 115, 1312

\bibitem[\protect\citeauthoryear{{Busso}, {Gallino} \& {Wasserburg}}{{Busso}
  et~al.}{1999}]{1999ARA&A..37..239B}
{Busso} M.,  {Gallino} R.,    {Wasserburg} G.~J.,  1999, ARA\&A, 37, 239

\bibitem[\protect\citeauthoryear{{Chieffi}, {Limongi} \& {Straniero}}{{Chieffi}
  et~al.}{1998}]{1998ApJ...502..737C}
{Chieffi} A.,  {Limongi} M.,    {Straniero} O.,  1998, ApJ, 502, 737

\bibitem[\protect\citeauthoryear{{Cristallo}, {Straniero}, {Lederer} \&
  {Aringer}}{{Cristallo} et~al.}{2007}]{2007ApJ...667..489C}
{Cristallo} S.,  {Straniero} O.,  {Lederer} M.~T.,    {Aringer} B.,  2007,
  ApJ, 667, 489

\bibitem[\protect\citeauthoryear{{Dulick}, {Bauschlicher} Jr., {Burrows},
  {Sharp}, {Ram} \& {Bernath}}{{Dulick} et~al.}{2003}]{2003ApJ...594..651D}
{Dulick} M.,  {Bauschlicher} Jr. C.~W.,  {Burrows} A.,  {Sharp} C.~M.,  {Ram}
  R.~S.,    {Bernath} P.,  2003, ApJ, 594, 651

\bibitem[\protect\citeauthoryear{{Ferguson}, {Alexander}, {Allard}, {Barman},
  {Bodnarik}, {Hauschildt}, {Heffner-Wong} \& {Tamanai}}{{Ferguson}
  et~al.}{2005}]{2005ApJ...623..585F}
{Ferguson} J.~W.,  {Alexander} D.~R.,  {Allard} F.,  {Barman} T.,  {Bodnarik}
  J.~G.,  {Hauschildt} P.~H.,  {Heffner-Wong} A.,    {Tamanai} A.,  2005, ApJ,
  623, 585

\bibitem[\protect\citeauthoryear{{Kupka}, {Ryabchikova}, {Piskunov}, {Stempels}
  \& {Weiss}}{{Kupka} et~al.}{2000}]{2000BaltA...9..590K}
{Kupka} F.~G.,  {Ryabchikova} T.~A.,  {Piskunov} N.~E.,  {Stempels} H.~C.,
  {Weiss} W.~W.,  2000, BaltA, 9, 590

\bibitem[\protect\citeauthoryear{{Lodders}}{{Lodders}}{2003}]{2003ApJ...591.12%
20L}
{Lodders} K.,  2003, ApJ, 591, 1220

\bibitem[\protect\citeauthoryear{{Marigo}}{{Marigo}}{2002}]{2002A&A...387..507%
M}
{Marigo} P.,  2002, A\&A, 387, 507

\bibitem[\protect\citeauthoryear{{Plez}, {van Eck}, {Jorissen}, {Edvardsson},
  {Eriksson} \& {Gustafsson}}{{Plez} et~al.}{2003}]{2003IAUS..210P..A2P}
{Plez} B.,  {van Eck} S.,  {Jorissen} A.,  {Edvardsson} B.,  {Eriksson} K.,
  {Gustafsson} B.,  2003, in {Piskunov} N.,  {Weiss} W.~W.,   {Gray} D.~F.,
  eds, Modelling of Stellar Atmospheres Vol.~210 of IAU Symposium, p.~2P

\bibitem[\protect\citeauthoryear{{Rogers} \& {Iglesias}}{{Rogers} \&
  {Iglesias}}{1993}]{1993npsp.conf..221R}
{Rogers} F.~J.,  {Iglesias} C.~A.,  1993, in {Nemec} J.~M.,  {Matthews} J.~M.,
  eds, IAU Colloq. 139, p.~221

\bibitem[\protect\citeauthoryear{{Seaton}}{{Seaton}}{2005}]{2005MNRAS.362L...1%
S}
{Seaton} M.~J.,  2005, MNRAS, 362, L1

\bibitem[\protect\citeauthoryear{{Straniero}, {Dom{\'{\i}}nguez}, {Cristallo}
  \& {Gallino}}{{Straniero} et~al.}{2003}]{2003PASA...20..389S}
{Straniero} O.,  {Dom{\'{\i}}nguez} I.,  {Cristallo} R.,    {Gallino} R.,
  2003, Publications of the Astronomical Society of Australia, 20, 389

\end{thebibliography}
\end{document}